\documentclass[preprint2]{emulateapj}

\usepackage{xspace}

\def \ergsec{\ensuremath{\mathrm{erg}\,\mathrm{s}^{-1}}}

\def \chandra{\emph{Chandra}\xspace} 
\def \xmm{\emph{XMM-Newton}\xspace}
\def \suzaku{\emph{Suzaku}\xspace}
\def \fe{Fe\xspace}
\def \ka{K$\alpha$\xspace}

\def \gcas{$\gamma$ Cas\xspace}

\shorttitle{\chandra and \suzaku observations of HD110432}
\shortauthors{Torrej\'on et al.}

\begin{document}


\title{ \chandra and \suzaku observations of the Be/X-ray star HD110432.}


\author{J.M. Torrej\'on\altaffilmark{1},
N.S. Schulz\altaffilmark{2}, M.A. Nowak\altaffilmark{2} }

\affil{$^{1}$Instituto de F\'isica Aplicada a las Ciencias y las Tecnolog\'ias, Universidad de Alicante, E03080 Alicante, Spain; jmt@ua.es} 

\affil{$^{2}$MIT Kavli Institute for Astrophysics and Space Research, Cambridge MA 02139, USA}




\begin{abstract}

We present an analysis of a pointed 141 ks \chandra high resolution transmission gratings observation of the Be X-ray emitting star HD110432, a prominent member of the $\gamma$ Cas analogs. This observation represents the first high resolution spectrum taken for this source as well as the longest uninterrupted observation of any $\gamma$ Cas analog. The \chandra lightcurve shows a high variability but its analysis fails to detect any coherent periodicity up to a frequency of 0.05 Hz. Hardness ratio versus intensity analyses demonstrate that the relative contributions of the [1.5-3] \AA, [3-6] \AA\ and [6-16] \AA\ energy bands to the total flux changes rapidly in the short term. The analysis of the \chandra HETG spectrum shows that, to correctly describe the spectrum, three model components are needed. Two of those components are optically thin thermal plasmas of different temperatures ($kT\approx$ 8--9 and 0.2--0.3 keV respectively) described by the models \texttt{vmekal} or \texttt{bvapec}. The Fe abundance in each of these two components appears equal within the errors and is sligthly subsolar with $Z\approx 0.75\,Z_{\odot}$. The \texttt{bvapec} model describes better the Fe L transitions although it can not fit well the Na \textsc{xi} Ly $\alpha$ line, at 10.02 \AA, which appears to be overabundant. Two different models seem to describe well the third component. One possibility is a third hot optically thin thermal plasma at $kT=$16--21 keV with an Fe abundance $Z\approx 0.3\,Z_{\odot}$, definitely smaller than for the other two thermal components. Furthermore, the \texttt{bvapec} model describes well the Fe K shell transitions because it accounts for the turbulence broadening the line Fe \textsc{xxv} and Fe \textsc{xxvi} lines with a $v_{turb}\approx1200$ km/s. These two lines, contributed mainly by the hot thermal plasma, are significantly wider thant the Fe K$\alpha$ line whose FWHM$<5$ mA is not resolved by \chandra. Alternatively, the third component can be described by a powerlaw with a photon index $\Gamma=1.56$. In either case, the \chandra HETG spectrum establishes that each one of these components must be modified by distinct absorption columns. The analysis of a non contemporaneous 25 ks \suzaku observation shows the presence of a hard tail extending up to at least 33 keV. The \suzaku spectrum is described with the sum of two components: an optically thin thermal plasma at $kT\approx 9$ keV and $Z\approx 0.74\,Z_{\odot}$; and a very hot second plasma with $kT\approx 33$ keV or, alternatively, a powerlaw with photon index $\Gamma=1.58$. In either case, each one of the two components must be affected by different absorption columns. Therefore, the $kT=$ 8--9 keV component is definitely needed while the nature of the harder emission can not be unambiguously established with the present data sets. The analysis of the Si \textsc{xiii} and S \textsc{xv} He like triplets present in the \chandra spectrum point to a very dense ($n_{e}\sim 10^{13}$ cm$^{-3}$) plasma located either close to the stellar surface ($r<3R_{*}$) of the Be star or, alternatively, very close ($r\sim 1.5R_{WD}$) to the surface of a (hypothetical) WD companion. We argue, however, that the available data supports the first scenario.

\end{abstract}


\section{Introduction}

Until recently, $\gamma$ Cas (B0.5IVe) stood alone as the major
exception to our understanding of the X-ray emission from massive
stars. Its X-ray spectrum is dominated by very hot ($kT\simeq$ 10 keV)
thermal emission at an X-ray luminosity of $\sim 10^{32}$
\ergsec in the 2-20 keV energy band, comparable to the brightest OB stars, much less than typical Be X-ray
binaries ($\sim 10^{35-37}$ ergs s$^{-1}$) and even below low luminosity ($\sim 10^{33-35}$ ergs s$^{-1}$) High Mass X-ray binaries like X Per \citep{palomb07}, SAX J1324.4$-$6200 and SAX J1452.8$-$5949 \citep{kaur09}. In $\gamma$ Cas the leading hypothesis for many years was accretion onto a degenerate companion, most probably a white dwarf
\citep{kubo,owens}. Evolutionary calculations have shown
that there should be a large number of binary systems with white dwarf
accretors \citep{waters,raguzova}.  These systems would show
far ultraviolet emission and should be observable as low luminosity
X-ray emitters ($L_{\rm X}\sim 10^{32}$ ergs s$^{-1}$). To date however, no actively accreting Be + White Dwarf
(Be + WD) system has been convincingly identified (see, however, Sturm et al. 2012). 

In the past years, a growing number of $\gamma$ Cas analogs have been
identified among late O and early B emission line stars, leading to the
definition of a new class of objects: HD110432
(B0.5IIIe; Torrej\'on \& Orr 2001, Lopes de Oliveira et al. 2007), HD119682 (B0.5Ve; Rakowski et al. 2006, Safi-Harb et al. 2007), HD 161103 (B0.5e), SAO49725 (B0.5e), NGC6649 (B1.5e), SS397 (B0.5e; Lopes de Oliveira et al. 2006), HD157832 (B1.5V; Lopes de Oliveira \& Motch 2011). As illustrated by
\cite{motch}, $\gamma$ Cas analogs share a very narrow
range of optical and X-ray properties: spectral types from B0.5 to B1.5,
slightly evolved, moderate and variable X-ray luminosities (of the order of
10$^{32}$ \ergsec in the 0.2 - 12 keV range) dominated by very hot ($\simeq$ 10 keV) thermal
plasmas, extended circumstellar disks  H$\alpha$ emissions with EW(H$\alpha$) $\sim 30-40$
\AA. None of them, except \gcas, have so far shown any
sign of binarity. \gcas is now known to be a wide binary. Its X-ray emission, however is devoid of any coherent modulation, orbital or otherwise.

The nature of this X-ray emission remains a mystery. Two mechanisms
have been proposed: in general, the broadband X-ray spectrum is reminiscent of
Cataclysmic Variables. The accretion onto a WD provides a
natural explanation for the hot thermal emission (as opposed to the
powerlaw displayed by NS systems in HMXB) while the smaller potential well of
the WD accounts naturally for the lower luminosity. This picture,
however, has some difficulties to fit into our current understanding of stellar
and binary evolution. Indeed, the progenitor of the WD should have been more
massive than the (early) Be star and it is by no means clear how a
massive star of 20$M_\odot$ can end as a WD. In the case of HD110432
no binary companion is known either. A tentative periodic
pulsation of 14 ks was claimed by \cite{to01} using a {\it BeppoSAX} observation. Later, however, \cite{lopes07} found no evidence of such a pulsation, nor any other, in an extensive analysis of three \xmm observations. On the other hand, \gcas is now known to be a wide binary with a $\sim 1 M_{\odot}$ star in a 203.59 d period orbit \citep{harmanec00}. This makes accretion very inefficient. According to current simulations \citep{okazaki02}, it is not possible to completely exclude the hypothesis in which matter is channeled in arm oscillations towards the (tentative) companion. Even in such a case, the WD would be more likely in a circular orbit truncating the disk, which renders the accretion mechanism more inefficient to explain the X-ray emission.

The alternative explanation is magnetic disc-Be star
interaction \citep{robinson02,smith04,smith06}. This model can explain not only the X-ray properties
but also the related UV and optical signatures, without the need to
introduce an unseen compact companion. The extremely hot X-ray
emitting gas is hypothesized to be entrained in a tangled magnetic field which produces also soft X-rays from
flare/disk-cloud interactions as well as cold fluorescence from
the disk. This interpretation, however, poses serious
challenges to the theory. A key ingredient is the presence of
a magnetic field, and it is by no means clear how such a field could
be produced and sustained in the radiative mantle of an OB star. Despite this, there is now a growing number of early type stars where the magnetic field has been detected and measured, for example $\theta^{1}$ Ori C \citep{donati02}, HD 108 \citep{martins10}, $\tau$ Sco \citep{donati06a} and HD191612 \citep{donati06b}.  On
the other hand, other Be stars, with similar spectral types and
comparably large disks, do not show any X-ray emission at all. In either case our
current understanding of stellar and binary structure
and evolution will be challenged. To characterize and understand the
class of $\gamma$ Cas analogs is, therefore, very important.

In this work we present an analysis of a \chandra gratings observation of the Be X-ray source HD110432 (1H1249$-$637 = BZ Cru, B0.5IIIe). \cite{to01} (hereinafter TO01) performed an analysis of {\it BeppoSAX} data describing the spectrum in terms of an optically thin thermal plasma at 10.55 keV. This was interpreted under the Be + WD scenario. The similarity of this system with \gcas was first noted by \cite{robinson02}. \cite{smith06}, performed an extensive analysis of optical data which supported the magnetically active star hypothesis, and definitely established the similarity with \gcas.  Later, \cite{lopes07} (hereinafter LO07), in an extensive analysis of \xmm PN and MOS data, at CCD resolution, found that the correct description of the system required several optically thin thermal plasmas of different temperatures and, perhaps, different absorption columns. The source spectrum was found to vary strongly in time. Again, the X-ray behavior is reminiscent of that of \gcas but the true nature of the object (whether single active Be star or Be binary) remains elusive. In order to shed new light on the nature of the system, we performed a 141 ks pointed observation of HD110432 with \chandra gratings. This observation represents the highest resolution X-ray spectrum ever taken for a $\gamma$ Cas analog as well as the longest uninterrupted lightcurve of HD110432 performed so far. In order to complement our understanding of the underlying broadband continuum we have also used a 25 ks \suzaku XIS and HXD observation taken from the public archive.

\section{Observations}

\begin{deluxetable}{lccc}
\tabletypesize{\scriptsize}
\tablecaption{Observations analyzed} 
\tablewidth{0pt}
\tablehead{
  \colhead{Telescope}  & \colhead{ObsID} & \colhead{Date} & \colhead{t$_{\rm exp}$} \\
  &   &   &  (ks)  \\
}
\startdata
\chandra  & 9947 &  2009-11-13 20:09:37  &  141.28  \\
         &      &    &  \\
\suzaku   & 403002010 & 2008-09-09 21:45:28 & 25.33 
\enddata

\label{tab:obs}
\end{deluxetable}

\subsection{\chandra-HETG data}

We have processed and analyzed the \chandra data from the ObsID 9947. The High Energy Transmission Gratings spectrometer (HETG; Canizares et al. 2005) acquired data during a total of 141.28 ks. Both High Energy Gratings (HEG; 1.55-17.7\AA) and Medium Energy Gratings (MEG; 1.55-31 \AA) were used for the analysis. After excluding wavelength ranges with insufficient signal, we obtained spectral coverage from 1.6 to 16 ~\AA. The spectra and the response files (arf and rmf)
were extracted using standard procedures with the CIAO software (v 4.4). The data were binned to
match the resolution of HEG (0.012 \AA\ {\it FWHM}) and MEG (0.023
\AA\ {\it FWHM}) and have a minimum S/N ratio of 7. First dispersion orders ($m=\pm 1$) were fitted
simultaneously. The peak count rate in a single (MEG) gratings arm spectrum is $\approx $ 0.04 cts/sec/\AA, which is well below the level at which pileup is a concern\footnote{See \emph{The Chandra ABC Guide to Pileup}, v.2.2, \texttt{http://cxc.harvard.edu/ciao/download/doc/pileup-abc.pdf}}.

The analysis was performed with the 
Interactive Spectral Interpretation System (ISIS) v 1.6.1-24 \citep{houck00}. For the analysis we use the abundance table from \citet{anders89} and cross sections
from \citet{balun92}.

\subsection{\suzaku data}

We have processed and analyzed public data from instruments on board \suzaku \citep{mitsu07}. We used standard procedures to reduce the X-ray Imaging Spectrometer (XIS; Koyama et al. 2007) CCD detector (0.3-10 keV) and the Hard X-ray Detector (HXD; Takahashi et al. 2007). The standard screening criteria (satellite out of the South Atlantic Anomaly, Earth elevation angle $> 5^{\rm o}$ and Day-Earth elevation angle $> 20^{\rm o}$) were applied to create the Good Time Intervals, GTI. Events were readout from either $3\times 3$ or $5\times 5$ pixel islands. We created individual spectra and response files for each detector and data mode combination. Response matrices and effective area files were created using the \texttt{xisrmfgen} and \texttt{xissimarfgen} tasks respectively. All three XIS detectors, XIS 0, 1 and 3, were combined and rebinned to have a minimum S/N ratio of 7 (XIS 2 was lost during a micrometeor hit in late 2006).

For the HXD, only data from the PIN diode detector (10-70 keV) presented enough signal to produce a useful spectrum. The PIN data were extracted from the cleaned event files in the \texttt{hxd/event-cl} directories. The source spectra exposure times were corrected with the \texttt{hxdcorr} tool. The PIN spectra were grouped to have a signal-to-noise ratio $\geq 10$ in each energy bin.

\section{Timing analysis}

\begin{figure*}
\includegraphics[angle=-90,width=2\columnwidth]{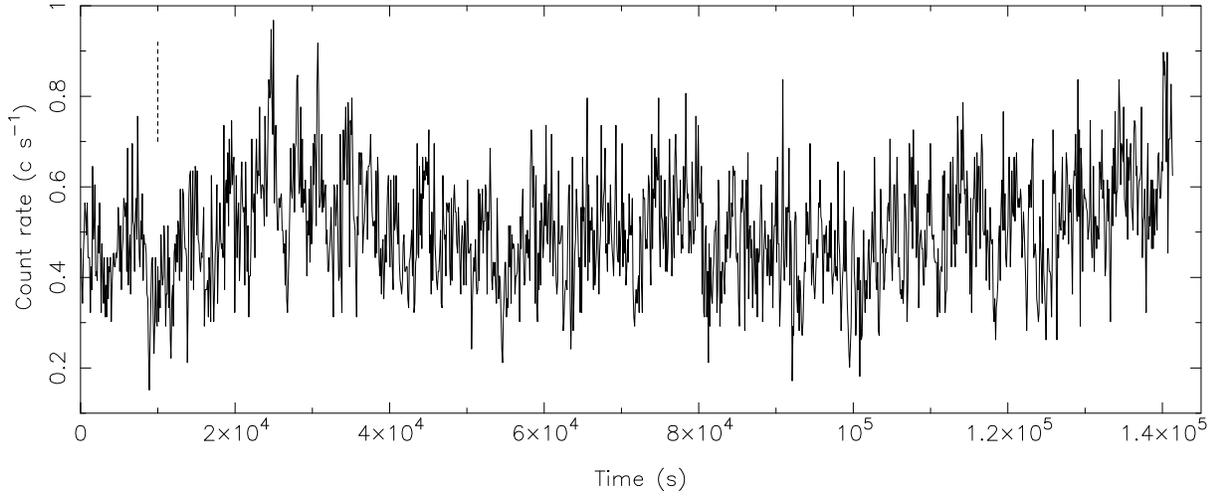}
\caption{\chandra 141 ks lightcurve, in the 1.6 - 16 \AA\ wavelength range, rebinned in 100 s intervals. The vertical dashed line in the upper left corner represents the typical error bar per bin.}
\label{fig:chandra_lc}
\end{figure*}

\begin{figure}
\includegraphics[angle=-90,width=\columnwidth]{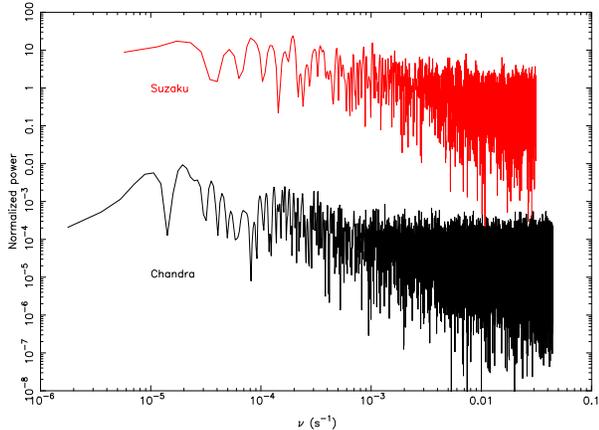}
\caption{Power spectrum corresponding to the 10 s binned \chandra lightcurve (bottom) and 16 s binned Suzaku lightcurve (top). No coherent period stands out. In particular, the 14 ks pulsation ($\nu=7.14\times 10^{-5}$ Hz) is not detected.}
\label{fig:chandra_powerspectrum}
\end{figure}

Our 141 ks \chandra observation represents the longest uninterrupted observation of this source (in fact, of any $\gamma$ Cas analog) made so far. This allows us to search for coherent pulsations over a wide range of frequencies. In Fig. \ref{fig:chandra_lc} we present the \chandra lightcurve binned to 100 s bins. Both HEG and MEG have been added and all orders have been used. As can be seen, the source is very variable on short timescales, showing a characteristic flaring pattern. In the long time scale, however, the source seems to lack of particularly big flares and looks quite stable. In order to search for coherent pulsations we have analyzed the lightcurve with a time resolution of 10 s\footnote{The HETG data appeared too noisy with smaller bins}. In Fig. \ref{fig:chandra_powerspectrum}, bottom plot, we show the normalized Fast Fourier Transform periodogram of the lightcurve, using the Lomb - Scargle technique. As can be seen, no significant period stands out. We have performed the same search using larger time bins to reduce errors. We fail to detect any significant period. \cite{to01} suggested the possibility of a 14 ks pulsation in the {\it BeppoSAX} lightcurve of HD110432. This detection, however, was rather uncertain because the observation covered just one cycle. Later, \cite{lopes07}, did not find any pulsation at such a period in any of the three longer \xmm EPIC lightcurves studied. In the present data set we do not find evidence of any significant period up to the Nyquist frequency (0.05 Hz). 

\begin{figure}
\includegraphics[angle=-90,width=\columnwidth]{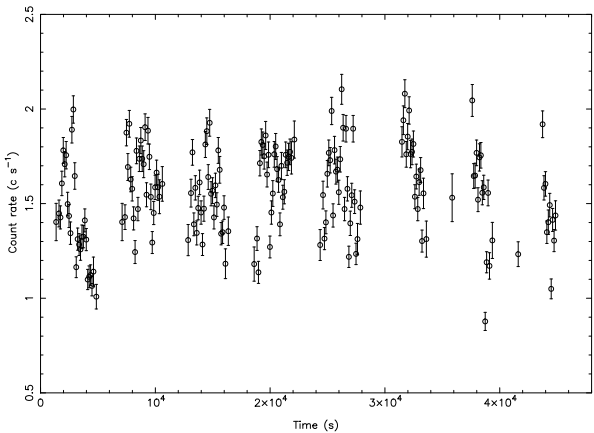}
\caption{\suzaku 20 ks lightcurve in the 0.5 - 9 keV band shown in time bins of 200 s. }
\label{fig:suzaku_lc}
\end{figure}

In Fig.\ref{fig:suzaku_lc} we show the \suzaku 0.5--9 keV XIS lightcurve in 200 s bins. The lightcurves shows, again, high variability on short timescales. The corresponding power spectrum is shown in Fig. \ref{fig:chandra_powerspectrum}, upper plot, where we have used 16 s time bins. As can be seen, no significant period stands out either. In summary, all attempts made so far to find a coherent peridod have failed.

\begin{figure}
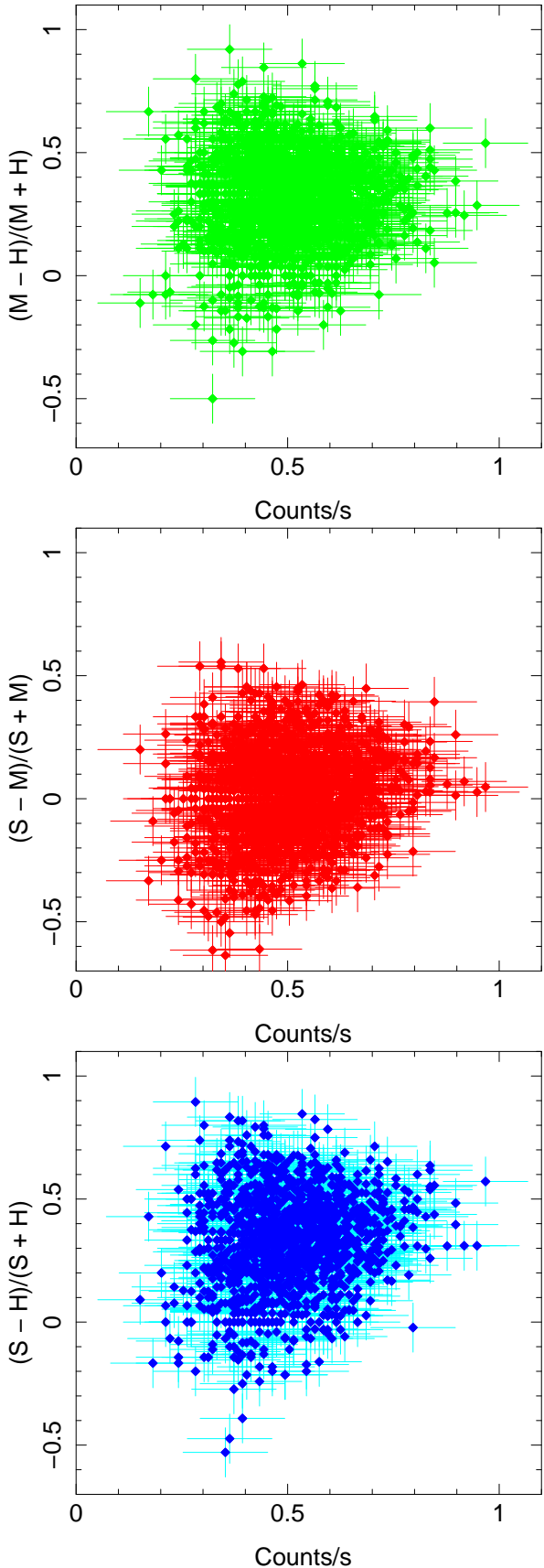

\includegraphics[angle=0,width=0.9\columnwidth]{colors_mh.ps}
\includegraphics[angle=0,width=0.9\columnwidth]{colors_sm.ps}
\includegraphics[angle=0,width=0.9\columnwidth]{colors_sh.ps}
\caption{\chandra color-color plots for several combination of hardness ratios: medium/High, soft/medium and soft/high.}
\label{fig:colors}
\end{figure}

\begin{figure}
\includegraphics[angle=0,width=0.9\columnwidth]{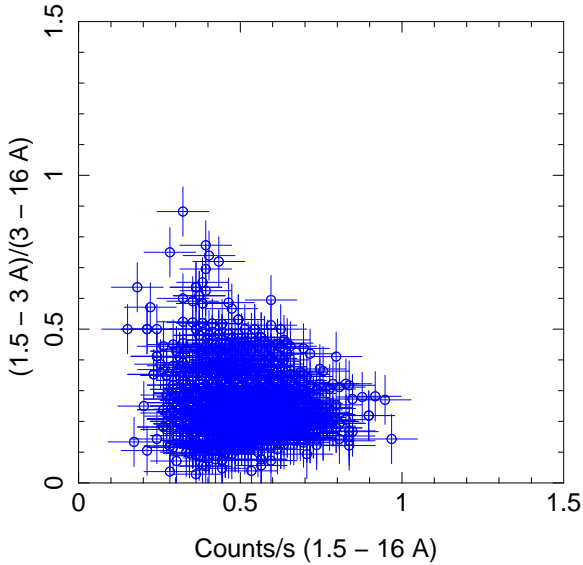}
\caption{\chandra color-magnitude diagram. Color is defined as the flux ratio between intervals (1.5 - 3 \AA)/(3 - 16 \AA).  }
\label{fig:chandra_color_mag}
\end{figure}

In order to explore the hardness variations of the X-ray source we have extracted the \chandra lightcurves in the wavelength intervals 1.5--3 \AA, 3--6 \AA\ and 6--16 \AA, referred to here as hard (H), medium (M) and soft (S) bands respectively. In Fig. \ref{fig:colors} we show the variation of medium/high, soft/medium and soft/high hardness ratios with counts. As can be seen, the source does not present any correlation of hardness ratio with source intensity in any of the energy bands. The points are scattered more or less uniformly around a mean value. The differences between hardness values are, however, significant.  In Fig. \ref{fig:chandra_color_mag} we show the ratio of the hard (1.5--3 \AA) to medium/soft energy band (3--16 \AA) with the intensity. This plot can be compared with Fig. 6 in LO07. Specifically, the behaviour during the \chandra\ observation seems to be very similar to the {\it XMM} observation 1 analyzed by LO07. This reflects the rapid variability of the X-ray energy distribution on short timescales. However, there seem to be no large variations on longer timescales. 

\section{Spectral analysis}

\subsection{Continuum and physical models}

\begin{deluxetable*}{lllll}
\tabletypesize{\scriptsize}
\tablecaption{Model parameters for \chandra\ data only. } 
\tablewidth{0pt}
\tablehead{
  \colhead{Component}  & \colhead{Parameter} &  & \colhead{Value}  &\\
}
\startdata
     &        & Model 1 & Model 2 & Model 3\\
     &        &    &   \\
edge & E (keV) & 7.125$^{\rm a}$ & 7.125$^{\rm a}$ & 7.125$^{\rm a}$\\
     & $\tau$ & 0.23$^{+0.15}_{-0.10}$ & 0.25$^{+0.13}_{-0.10}$ & 0.19$^{+0.14}_{-0.11}$ \\
     &        &    &  & \\
powerlaw & $N^{\rm b}_{\rm H}$  & 3.04$^{+0.15}_{-0.04}$ & ... & ...\\
         & Norm & 0.005$^{+0.002}_{-0.001}$ & ...& ...\\
         & $\Gamma$ & 1.56$^{+0.005}_{-0.10}$ & ...& ...\\
         &          &     &  & \\
Thermal (hot) & $N_{\rm H}$  & ... & 2.09$\pm$ 0.29 & 2.25$\pm{0.12}$\\
            & Norm &  ... & 0.023$\pm$0.002 & 0.024$\pm{0.004}$\\
            & $kT$ (keV) & ... & 22$^{+8}_{-5}$ & 16$^{+5}_{-2}$\\
            & $Z_{Fe}$  &  ... & 0.30$\pm$0.12 & 0.32$\pm$0.15\\
            & $v_{turb}$ (km/s) & ...  & ... & 1200$\pm{300}$\\
            &        &      &   & \\
Thermal (warm) & $N_{\rm H}$  & 0.89$\pm{0.10}$ & 0.72$\pm{0.12}$ & 0.70$\pm{0.13}$ \\
            & Norm & 0.017$^{+0.006}_{-0.002}$ & 0.0104$^{+0.0010}_{-0.0005}$ & 0.011$\pm{0.0003}$\\
            & $kT$ (keV) & 8.82$\pm{0.75}$ & 8.07$^{+0.80}_{-0.88}$ & 7.11$\pm{0.83}$ \\
            & $Z_{Fe}$  &  0.68$\pm{0.15}$  & 0.76$\pm$0.20 & 0.75$\pm{0.13}$\\
            &        &      &   & \\
Thermal (cool) & $N_{\rm H}$  & 0.17$\pm{0.05}$ & 0.15$^{\rm c}$ & 0.15$^{\rm c}$\\
            & Norm & 0.0006$\pm{0.0002}$ & 0.0006$\pm 0.0005$ & 0.00017$\pm{0.0005}$\\
            & $kT$ (keV) & 0.20$\pm{0.08}$ & 0.19$\pm$0.03 & 0.28$\pm{0.07}$\\
            & $Z_{Fe}$  &  0.68$\pm{0.15}$ & 0.76$\pm$0.20 & 0.75$\pm{0.13}$\\
            &        &      &   & \\
            & Flux$^{\rm d}$   &  3.78    & 3.78  & 3.78 \\
Fe K$\alpha$ & $\lambda$ & 1.9365$^{+0.0005}_{-0.0004}$ & 1.9366$^{+0.0005}_{-0.0004}$ & 1.9366$^{+0.0005}_{-0.0004}$ \\
             & Flux$^{\rm d}$ & 3.72$\pm{0.55}$ & 4.14$\pm{0.22}$ & 4.23$\pm{0.23}$\\
             & $FWHM$ & $\leq$ 5mA$^{\rm a}$ & $\leq$ 5mA$^{\rm a}$ & $\leq$ 5mA$^{\rm a}$\\
EW(eV/m\AA)     &        & 88.76/26.84  & 83.92/25.34  & 89.32/27.01\\
             &        &      &   & \\
 &  $\chi^{2}_{\rm r}$ (dof)  & 1.07 (932)   & 1.07 (932) & 1.05 (932)    
\enddata
\tablenotetext{a}{fixed}\tablenotetext{b}{All $N_{\rm}$ in units of $\times 10^{22}$ cm$^{-2}$}\\
\tablenotetext{c}{fixed at the interstellar value}
\tablenotetext{d}{Unabsorbed total flux in un its of $\times 10^{-11}$ erg s$^{-1}$ cm$^{-2}$}
\tablenotetext{}{Model 1: PO+2 VMEKAL, Model 2: 3 VMEKAL, Model 3: 3 BVAPEC}
\label{tab:results}
\end{deluxetable*}


\begin{figure*}
\includegraphics[angle=-90,width=2\columnwidth]{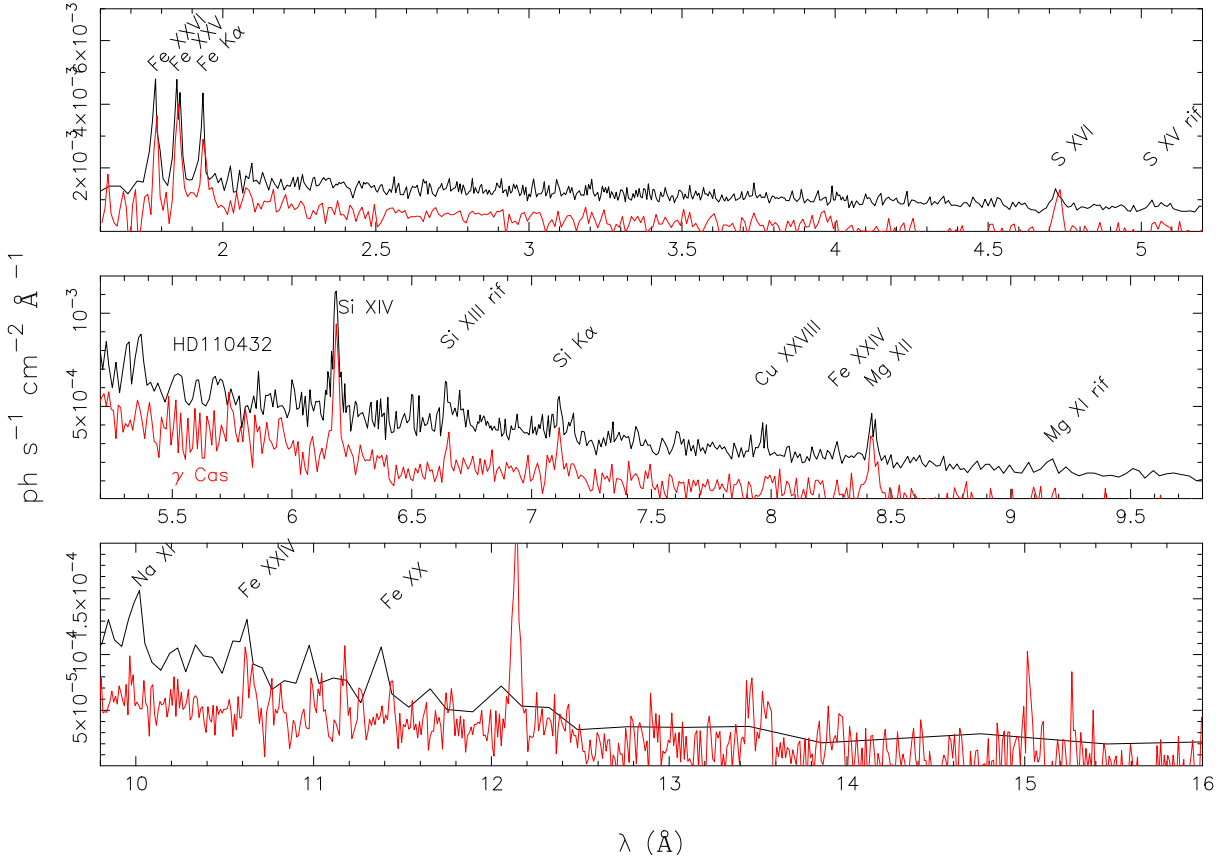}
\caption{The \chandra {\it HETG} unfolded spectrum of HD110432 (top, black) analyzed in this work. The HETG spectrum of \gcas, presented in Smith et al. (2004), is also shown (bottom, red) scaled and offset for comparison.}
\label{fig:nolines}
\end{figure*}

In previous works on HD110432 (LO07) and \gcas \citep{smith04, lopes10} it was shown that the description of the spectrum of these sources requires several optically thin thermal plasma components (\textsc{mekal, apec}) to correctly account for both the shape of the continuum and the emission line strengths. One of the critical points left open by LO07 was whether the different plasmas in HD110432 were affected by distinct absorption columns. The high resolution \chandra\ HETG spectrum will allow us to further investigate this issue. In Fig. \ref{fig:nolines} we show the \chandra\ {\it HETG} unfolded spectrum of HD110432.  

It has been shown in \cite{torr10}, that the Fe K fluorescence is always accompanied by the presence of a K edge in massive binaries. This edge is produced by those photons energetic enough to ionize the innermost K shell electrons of the atom. This hole is subsequently replenished by an electron from the upper level producing the observed fluorescence line. Consequently the Fe K edge is deeper than what is imposed by the curvature of the spectrum at low energies. In general, then, all the models will require an additional edge at 1.74 \AA\ (= 7.125 keV). We will fix the wavelength and let the optical depth $\tau$ vary free. Likewise, all models require the addition of a fluorescence line at $1.9366$ \AA, compatible with neutral Fe (\textsc{i} to \textsc{viii}) from the illumination of cold material or reflection onto the tentative WD. This line is not included in the plasma models.

As can be seen in Table \ref{tab:lines}, the emission lines present in the \chandra spectrum, trace a range of plasma temperatures. We have tried a single thermal component modified  by a single absorption column, using the \texttt{vmekal} model. As expected, this model does not produce a good fit to the continuum especially at low energies. The reduced $\chi_{r}^{2}=1.46$ for 938 degrees of freedom (dof). In order to account for the soft excess, and in agreement with LO07, we find that at least three plasmas, at different temperatures, are required. However, using a single absorption column to modify all three thermal components yields a reduced $\chi_{r}^{2}=1.25$ for 935 dof. The most evident discrepancies are still in the low energy continuum. In order to properly describe the whole {\it continuum} we find that three distinct absorption columns must be used to modify each of the three \texttt{vmekal} components. The fit improves significantly. The reduced chi square is now $\chi_{r}^{2}=1.19$ for 933 dof. An $F$ test gives an statistic value of $F=23.5$ and a probability for the addition of the individual absorptions not being significant is only $1.08\times 10^{-10}$. At this point the continuum is fairly well described both at high and at low energies by three optically thin thermal plasmas at $kT\approx22$, 8 and 0.2 keV respectively. Following previous works, we will call them {\it hot}, {\it warm} and {\it cool}  plasmas respectively. The overall Fe abundance is 0.48 solar. The main discrepancies are now in the line strenghts. In particular, it is evident that the Fe L shell transitions (Fe \textsc{xxiv}, Fe \textsc{xx}), that dominate the low energy spectral range, are severely underestimated by this model. Therefore, we have next investigated the elemental abundances in the the several plasmas by letting the Fe abundance to vary independently. We find that the abundance of the hot component is subsolar while the abundances of the warm and cool components are compatible between them. Therefore we tie the abundances in the warm and cool components and let the abundance in the hot component vary freely. The best fit parameters are shown in Table \ref{tab:results}, under Model 2 column. Owing to the normalization of the several components, the hot plasma dominates the spectrum, followed by the warm and cool plasmas respectively. The absorption columns follow this hierarchy as well. The absorption for the cool component turns out to be very low. We have fixed it to the ISM value ($N_{H}\approx 0.15\times 10^{22}$ cm$^{-2}$). The Fe abundance is subsolar. However, the Fe abundance of the hot component ($Z_{Fe}=0.30Z_{\odot}$) is significantly lower than for the warm and cool components ($Z_{Fe}=0.76Z_{\odot}$) which turn out to be equal within the errors and have been tied during the fittings.

\begin{figure*}
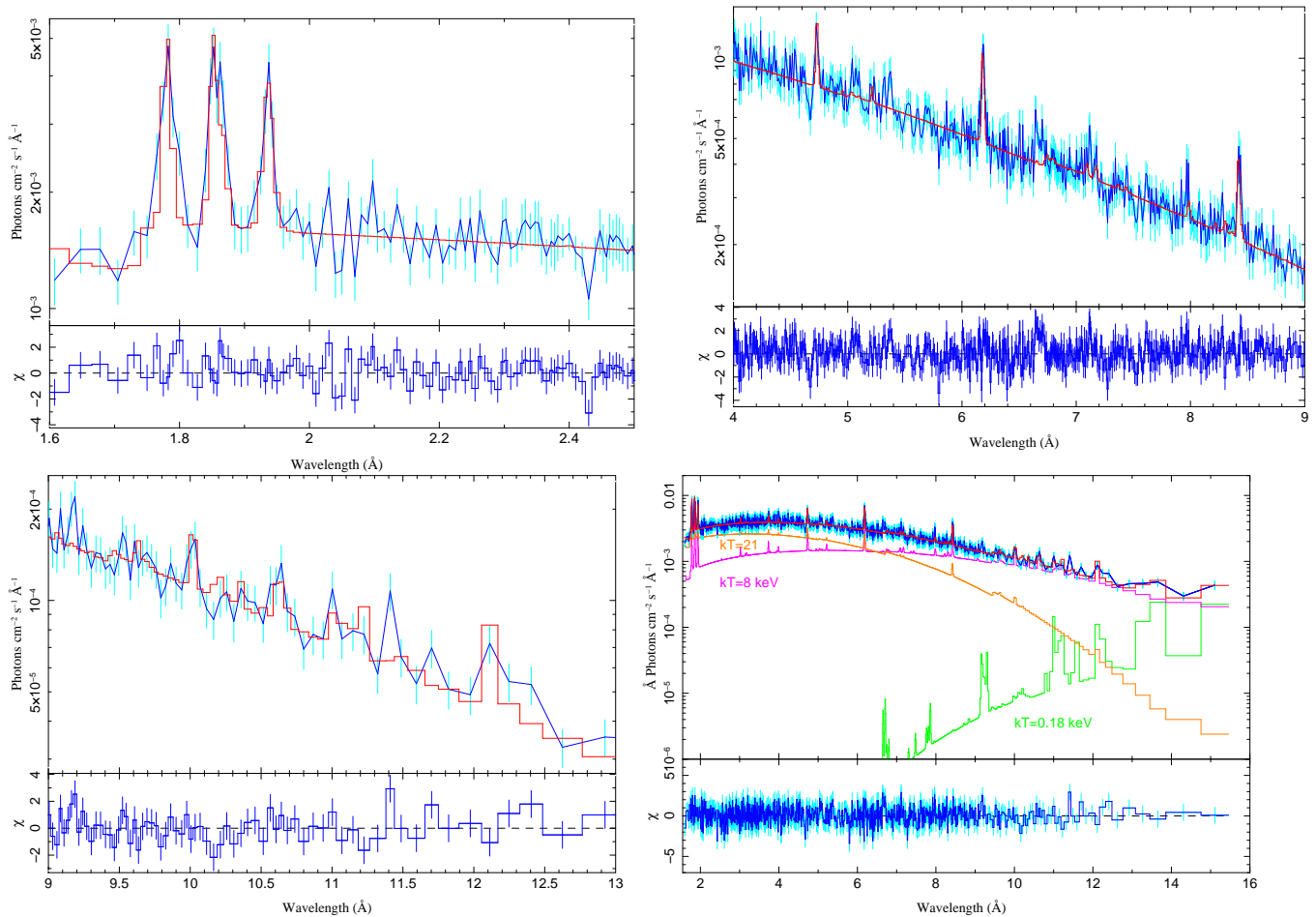

\includegraphics[angle=-90,width=\columnwidth]{unfold_1_2.ps}
\includegraphics[angle=-90,width=\columnwidth]{unfold_4_9.ps}
\includegraphics[angle=-90,width=\columnwidth]{unfold_9_13.ps}
\includegraphics[angle=-90,width=\columnwidth]{vmekal_unfold_components.ps}
\caption{The \chandra\ {\it HETG} unfolded spectrum of HD110432 with the \texttt{vmekal} model fit in the wavelength intervals 1.5-2.5 \AA, 4-9 \AA\ and 9-13 \AA. The fit in the Fe line complex region is not as good as with the colisional plasma model. The exact values of the plasma temperatures are quoted in Table \ref{tab:results}.}
\label{fig:vmekal}
\end{figure*}

In Fig. \ref{fig:vmekal} we plot the unfolded spectrum with the three \texttt{vmekal} composite model. In the first three panels we show the resulting fit for the continuum and the lines while in the fourth panel we show the individual contributions of each component. The reduced $\chi_{r}^{2}=1.07$ for 932 dof. As can be seen, the continuum is well accounted for while some discrepancies still exist in the lines. Two cases stand out. First the observed strength of the Na \textsc{xi} Ly$\alpha$ line at 10.02 \AA\ is much stronger than predicted by the model using solar abundance. We let it vary freely, keeping it tied for the three components. The resulting Na abundance is very high ([$\sim 15\pm 5$]$Z_{\odot}$). Unfortunately the spectrum starts to be noisy at these wavelengths but a high Na abundance seems to be necessary to correctly fit the line. Second, the peak emission of the Fe \textsc{xxvi} Ly$\alpha$ and Fe \textsc{xxvi} lines, at 1.78 and 1.86 \AA\ respectively, seem slightly overpredicted while the observed profiles are definitely broader than the model. Therefore, the next step is to modify the width of the observed lines with the effects of turbulent velocity. For that purpose we use the {\texttt{bvapec} model, for collisionaly ionized plasma based on the \emph{Astrophysical Plasma Emission Code} (\textsc{apec}). 

In Table \ref{tab:results}, under Model 3, we show the corresponding fit parameters. The
hierarchy of normalizations and absorptions is still observed. The temperatures of the hot and warm components are lower, $kT=16.3$ and 7.1 keV respectively, while the temperature of the cool component is a bit higher, $kT=0.28$ keV, although its norm is smaller than in the previous model.

\begin{figure*}
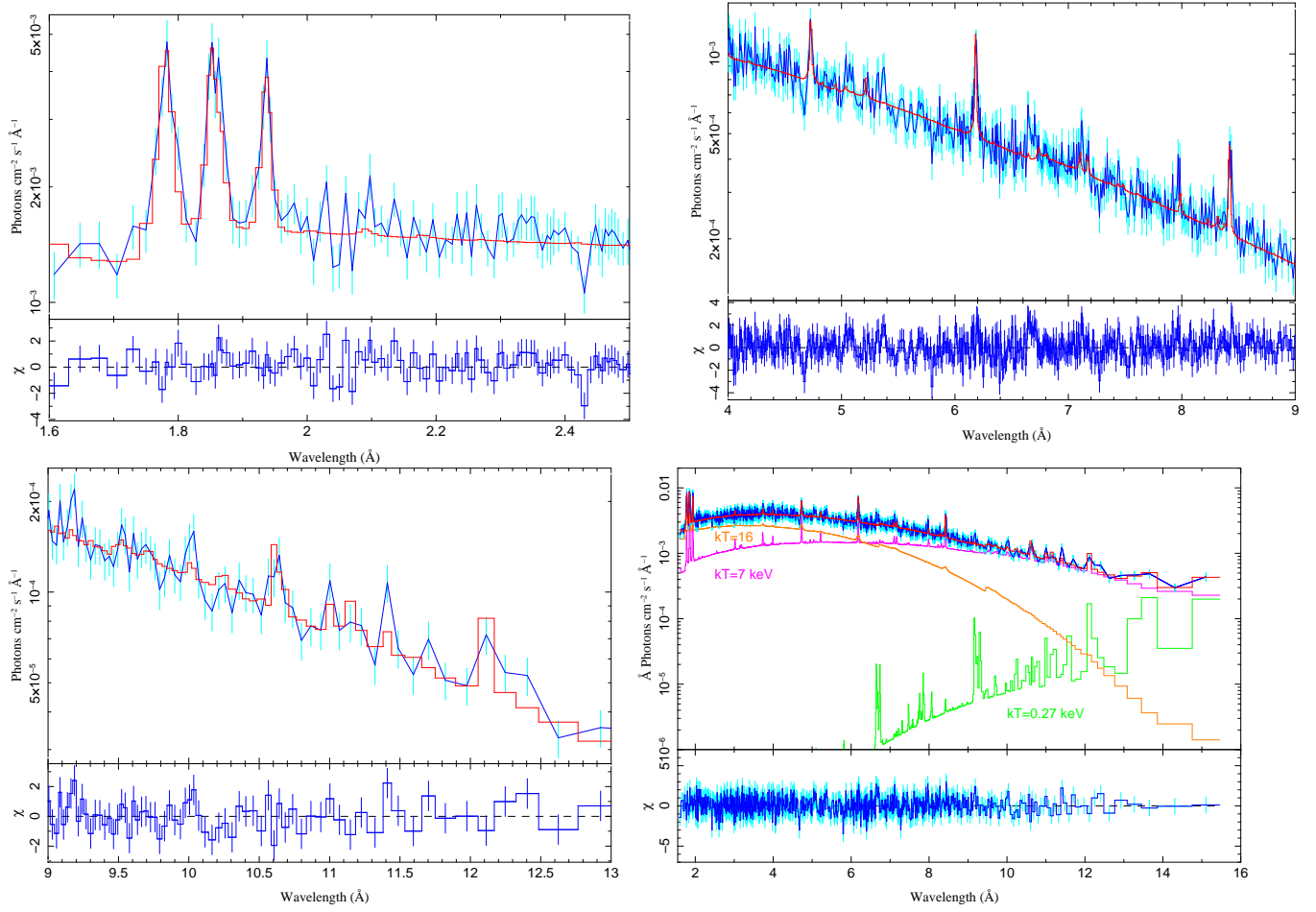

\includegraphics[angle=-90,width=\columnwidth]{bvapec_1_2.ps}
\includegraphics[angle=-90,width=\columnwidth]{bvapec_4_9.ps}
\includegraphics[angle=-90,width=\columnwidth]{bvapec_9_13.ps}
\includegraphics[angle=-90,width=\columnwidth]{bvapec_unfold_components.ps}
\caption{The \chandra\ {\it HETG} unfolded spectrum of HD110432 with the \texttt{bvapec} model fit in the wavelength intervals 1.5-2.5 \AA, 4-9 \AA\ and 9-13 \AA. The fit in the Fe line complex region is the best as it accounts simultaneously for the abundance and $v_{turb}$ effects. In the low energy tail the Na \textsc{xi} line at 10.02 \AA\ is not fit properly since the abundance of this element can not be set within this model. The exact values of the plasma temperatures are quoted in Table \ref{tab:results}.}
\label{fig:bvapec}
\end{figure*}

In Fig. \ref{fig:bvapec} we show the unfolded spectrum for this model. As can be seen, both  Fe \textsc{xxvi} and  Fe \textsc{xxv} are much better described now. These two lines are the main contributors, within the investigated spectral range, to the $v_{turb}$ determination. The required $v_{turb}$ is of the order of 1200 km/s. No turbulent broadening is needed to modify the warm and cool components. Note however, that the Mg (9.2 \AA) and Na (10.02 \AA) lines are underestimated. The \texttt{bvapec} model does not have the Na abundance as variable and then it cannot fit it well. This reinforces the conclusion from the previous analysis that the Na seems be highly abundant. However, the reduced $\chi_{r}^{2}=1.04$ for 931 dof is the best we have found in our analysis. It is plausible that a correct description of the whole spectrum will be a hybrid collisional/photoionized plasma. In the last panel of Fig. \ref{fig:bvapec} we show the relative contribution of the several plasmas. The prominent S\textsc{xvi}, Si\textsc{xiv} and Mg\textsc{xii} lines as well as the Fe L shell transitions are contributed mainly by the $kT=(8-9)$ keV plasma. Although it contributes also to Fe \textsc{xxv} and Fe \textsc{xxv} lines, these are dominated by the hot component. Finally, the $kT=0.3$ keV cool plasma contributes significantly only beyond $\lambda\sim 12$ \AA.

Given the limited \chandra bandpass, it is necessary to check the consistency of the hard tail with a hot (thermal or no-thermal) emission. Indeed at such high temperatures, a thin plasma displays already a powerlaw shape at the energies considered here. To that end we substitute the hot component by an absorbed powerlaw. The fit is also good and the parameters are given in Table \ref{tab:results}, under Model 1. The most significant change is the increase in the norm of the warm component which is now dominant. The main discrepancy between the data and the model is, again, in the line strengths. The Fe \textsc{xxv} (1.85 \AA) is overestimated while Fe \textsc{xx} (11.4 \AA) is underestimated. In terms of the $\chi_{r}^{2}$ and fit to the continuum, however, this model works equally well.

In order to try to break down this degeneracy we have tried to establish the underlying broadband continuum using \suzaku. In Fig. \ref{fig:suzaku_3aped} we present the combined XIS and HXD \suzaku data fitted with a three optically thin thermal plasmas. Significant emission is seen up to, at least, 33.5 keV even though the number of energy bins is very low at high energies. In the {\it BeppoSAX} data presented in TO01 this hard tail could be seen up to 50 keV. The S/N ratio of the PDS data on board BeppoSAX was, unfortunately, also low. However, a hard tail is definitely established for this system. Consistently with the previous discussion, the \suzaku data can be modeled with either a sum of thermal plasmas or a thermal-powerlaw mixture. The \suzaku data, however, do not require the cooler component. In Table \ref{tab:results_2} we present the details of the fit to the \suzaku data only. As can be seen in Tables \ref{tab:results} (Models 2 and 3) and \ref{tab:results_2} (Model 2), for pure thermal models, the source exhibits a complex variability, a fact already noted by LO07. The hot component is hotter during the \suzaku observation while the temperature of the warm component is similar. However, the norm of the hot and warm component was lower and higher respectively during the \suzaku observation. In the case of the thermal plasma + powerlaw model the most noticeable change is in the norm of the powerlaw which again was less important here that during the \chandra observation. This precludes any attempt to obtain physical parameters based on the joint fit to both datasets. Unfortunately, the low S/N ratio of the \suzaku HXD data does not allow us to determine its full extension. The goodness of the fit is comparably good for both models. We can not constrain unambiguously the very nature of the hard component with the available datasets and further observations will be required to establish this point.

\begin{deluxetable*}{llll}
\tabletypesize{\scriptsize}
\tablecaption{Model parameters for \suzaku\ data only.  } 
\tablewidth{0pt}
\tablehead{
  \colhead{Component}  & \colhead{Parameter} &  & \colhead{Value} \\
}
\startdata
     &        & Model 1 & Model 2 \\
     &        &         &      \\
Powerlaw    & $N_{\rm H}$($\times 10^{22}$ cm$^{-2}$) & 2.94$\pm$0.23& ... \\
         & Norm & 0.0036$\pm$ 0.0009 & ...\\
         & $\Gamma$ & 1.58$_{-0.02}^{+0.10}$ & ...\\
         &          &     &  \\
Thermal (hot) & $N_{\rm H}$($\times 10^{22}$ cm$^{-2}$) & ... & 2.12$\pm$0.23 \\    
            & Norm &  ...  & 0.015$\pm$ 0.005\\
            & $kT$ (keV) & ... & 33$^{+3}_{-5}$\\
            & Abund  &  ... & 0.20$\pm$ 0.25 \\
            &        &      &   \\
Thermal (warm) & $N_{\rm H}$($\times 10^{22}$ cm$^{-2}$) & 0.47$\pm$0.22 & 0.44$\pm$0.23 \\
            & Norm & 0.018$\pm$0.001 & 0.017$\pm$0.001 \\
            & $kT$ (keV) & 9.44$\pm$0.12 & 8.69$\pm$0.13  \\
            & Abund  &  0.73$\pm$0.23  & 0.71$\pm$0.25 \\
            &        &      &   \\
            & Flux$^{\rm d}$ &  3.74$\pm$0.13 &  3.75$\pm$0.13  \\
            &        &      &   \\
Fe K$\alpha$ & $\lambda$ & 1.938$\pm$0.002 & 1.938$\pm$0.002 \\
             & $FWHM$ & $\leq$ 5mA$^{\rm a}$ & $\leq$ 5mA$^{\rm a}$ \\
             & $EW$ (m\AA) & 29.20  &   30.2     \\
             & Flux$^{\rm e}$ & 4.26 & 3.58 \\
     &        &    &   \\
 & $\chi^{2}_{\rm r}$ (dof)   & 1.304 (190)  & 1.295 (189)    
\enddata
\tablenotetext{a}{fixed}\tablenotetext{b}{All $N_{\rm}$ in units of $\times 10^{22}$ cm$^{-2}$}\\
\tablenotetext{c}{fixed at the interstellar value}
\tablenotetext{d}{$\times 10^{-11}$ erg s$^{-1}$ cm$^{-2}$; [0.78-8.27] keV = [1.5-16] \AA}
\tablenotetext{e}{$\times 10^{-5}$ Photons s$^{-1}$ cm$^{-2}$}
\tablenotetext{}{Model 1: po+Mekal, Model 2: 2 Mekal}
\label{tab:results_2}
\end{deluxetable*}

\begin{figure}
\includegraphics[angle=-90,width=\columnwidth]{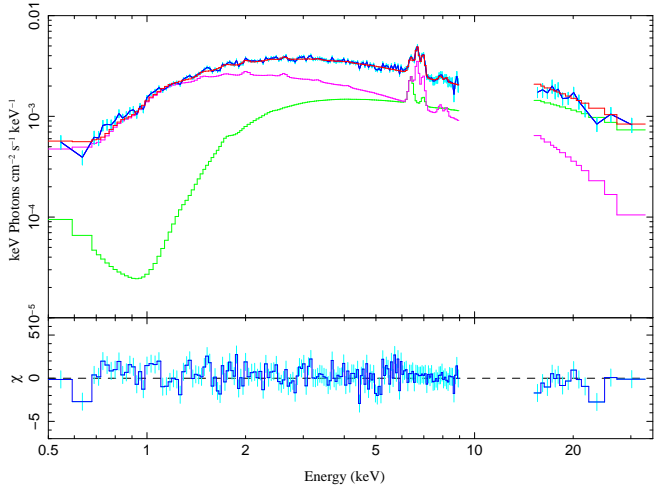}
\caption{\suzaku XIS and HXD spectra of HD110432 fitted with a 2 thin thermal plasmas modeled with \texttt{vmekal}. The parameters for two \texttt{bvapec} components are very similar.}
\label{fig:suzaku_3aped}
\end{figure}


\subsection{Cooling flow model}

\begin{deluxetable}{lll}
\tabletypesize{\scriptsize}
\tablecaption{Model parameters for the cooling flow model fit to \chandra\ data.  } 
\tablewidth{0pt}
\tablehead{
  \colhead{Component}  & \colhead{Parameter} &  \colhead{Value} \\
}
\startdata
     &        &            \\
edge & E (keV) & 7.125$^{\rm a}$ \\
     & $\tau$ & 0.42$\pm$ 0.02 \\
     &        &      \\
Powerlaw     & $N_{\rm H}$($\times 10^{22}$ cm$^{-2}$) & 1.18$\pm$0.23\\
     & Norm & 0.0026$\pm$ 0.0009 \\
         & $\Gamma$ & 1.13$_{-0.02}^{+0.10}$ \\
         &          &      \\
Thermal  & Norm &  0.027$\pm$ 0.003\\
         & $\alpha$ & 0.82$\pm$0.10   \\
            & $kT_{max}$ (keV) & 37.35$\pm$ 0.52 \\
            & Abund  &  1.26$\pm$ 0.23  \\
     &        &      \\
Fe K$\alpha$ & $\lambda$ & 1.938$\pm$0.002  \\
             & Flux$^{\rm d}$ & 3.74 \\
             & $FWHM$ & $\leq$ 5mA$^{\rm a}$ \\
             & $EW$ (m\AA) & 21.15   \\
     &        &     \\
     & $\chi^{2}_{\rm r}$ (dof)   & 1.24 (936)     
\enddata
\tablenotetext{a}{fixed}\tablenotetext{b}{All $N_{\rm}$ in units of $\times 10^{22}$ cm$^{-2}$}\\
\tablenotetext{d}{$\times 10^{-5}$ Photons s$^{-1}$ cm$^{-2}$}
\tablenotetext{}{Model: POWERLAW + CEMEKL}
\label{tab:results_cemekl}
\end{deluxetable}

One of the fingerprints of accretion on to the white dwarf (WD) in some cataclysmic variables is through the cooling of the post-shock X-ray plasma (see, for example, Mukai et al. 2003). In order to test this scenario we have tried to fit the cooling flow model \texttt{cemekl} to the \chandra data. A single absorbed \texttt{cemekl} component does not fit the data at all. An acceptable fit is obtained with the adition of a powerlaw. The best fit parameters are displayed in Table \ref{tab:results_cemekl}. As can be seen the maximum temperature displayed by the plasma is very high ($\sim 37$ keV) and the $\alpha$ parameter is close the adiabatic case ($\alpha=1$). This parameters are consistent with those found by LO07 for the same model. The photon index of the powerlaw is identical. This model, however, gives a much higher $\chi_{r}^{2}=1.24$ for 936 dof, than the models discussed in the previous section. The main discrepancies here are in the fit to the low energy continuum as well as the Fe \textsc{xxvi} Ly$\alpha$ line which is overestimated. We tried also other combinations like \texttt{cemekl}+\texttt{mekal} but the quality of the fit does not improve over the one presented here.

\subsection{Emission lines}



\begin{deluxetable*}{lcccccc}
\tabletypesize{\scriptsize}
\tablecaption{Emission lines identified in the \chandra spectrum of HD110432.}
\tablewidth{0pt}
\tablehead{
\colhead{Ion}  & \colhead{$\lambda$} & \colhead{Flux} & \colhead{$\sigma$} & \colhead{$EW$} & $kT_{\rm max}$ & \colhead{$EM$} \\
& & $\times 10^{-6}$ & & &  & $\times 10^{55}$\\
 & (\AA) &  (ph s$^{-1}$ cm$^{-2}$) & (\AA) & (m\AA) & (keV) & (cm$^{-3}$)\\
}
\startdata
Fe \textsc{xxvi} Ly$\alpha$ &  1.7821$\pm$ 0.0020 & 71.09$\pm$14 & 0.0098$\pm$ 0.0032 & 44.94$\pm$8.86  & 10.85 & 6.42$\pm$ 1.26\\
Fe \textsc{xxv} & 1.8550$\pm$ 0.0025  &75.90$\pm$12  &0.0088$\pm$ 0.0030  &48.42$\pm$7.70  & 5.43 & 5.98$\pm$0.92 \\
\fe \ka & 1.9366$_{-0.004}^{+0.005}$ & 41.95$_{-0.30}^{+1.20}$ & 0.005 & 27.09$\pm$0.33 & - & -\\
S \textsc{xvi} Ly$\alpha$ & 4.7287$\pm$ 0.0015  &16.04$_{-0.5}^{+0.7}$  &0.010$\pm$ 0.0020 & 20.49$\pm$0.77  & 2.17 & 1.62$\pm$0.06\\
& & & & & &  \\
S \textsc{xv} $r$ & 5.0387\tablenotemark{a}  & 5.10$\pm{2.48}$  &0.005 & 5.58$\pm$2.71  & 1.37 & 0.14$\pm$0.07 \\
S \textsc{xv} $i$ & 5.0665\tablenotemark{a}  & 3.73$\pm{1.53}$  &0.005 & 4.62 & - & -\\
S \textsc{xv} $f$ & 5.1015\tablenotemark{a}  & 3.21$\pm{1.55}$ &0.005 & 3.49 & - & -\\
& & & & & &  \\
Si \textsc{xiv} Ly$\alpha$ & 6.1856$_{-0.022}^{+0.021}$  & 19.78$_{-0.10}^{+0.45}$ & 0.0107$\pm$ 0.0020  & 41.39$\pm$0.76 & 1.37 & 0.75$\pm$0.01 \\
& & & & & & \\
Si \textsc{xiii} $r$ & 6.6479 &  4.88$\pm{1.74}$ & 0.005 & 12.67$\pm$4.52 & 0.86 & 0.06$\pm$0.02 \\
Si \textsc{xiii} $i$ & 6.6850 &  2.50$\pm{1.55}$ & 0.005 & 7.06  & - & -\\
Si \textsc{xiii} $f$ & 6.7395&  1.04$\pm{0.48}$ & 0.005 & 1.86 & - & -\\
& & & & & &  \\
Si \ka & 7.1186$_{-0.0010}^{+0.0001}$  & 4.26$\pm{0.20}$ & 0.005 & 12.02$\pm$0.56  & 0.86 & - \\
Cu \textsc{xxviii} & 7.9667$_{-0.009}^{+0.008}$ & 2.18$_{-0.07}^{+0.21}$ & 0.005 & 8.82$\pm$0.57 &  & \\
Fe \textsc{xxiv} & 8.2783$_{-0.007}^{+0.008}$ & 2.04$_{-0.05}^{+0.15}$ & 0.005 & 9.43$\pm$0.46  & 1.72 & 0.02$\pm$0.001\\
Mg \textsc{xii} Ly$\alpha$ & 8.4279$\pm$ 0.0032  & 8.42$_{-0.30}^{+0.17}$ & 0.015$_{-0.0002}^{+0.001}$ & 41.74$\pm$1.14  & 0.86 & 0.24$\pm$0.007 \\
& & & & & & \\
Mg \textsc{xi} $r$ & 9.1687 & 2.17$\pm{1.44}$& 0.005 & 16.74$\pm$10 & 0.54 & 0.03$\pm$0.01\\
Mg \textsc{xi} $i$ & 9.2300 & 0.52$\pm{0.25}$& 0.005 & 6.13$\pm$2.92 & - & -\\
Mg \textsc{xi} $f$ & 9.3136 & 0 & 0.005 & 0  & - & -\\
& & & & & & \\
Na \textsc{xi} Ly$\alpha$ & 10.0214$_{-0.019}^{+0.033}$  &2.50$_{-0.22}^{+0.22}$ & 0.005  & 22.42$\pm$1.79   & 0.69 & 1.18$\pm$0.09\\
Fe \textsc{xxiv} & 10.6279$_{-0.022}^{+0.035}$ & 2.53$_{-0.20}^{+0.17}$ & 0.005 & 28.92$\pm$2.06  & 1.72 & 0.02$\pm$0.001 \\
Fe \textsc{xx} & 11.4237$_{-0.020}^{+0.043}$ & 2.62$_{-0.18}^{+0.17}$ & 0.005 & 40.13$\pm$2.76 & 0.86 & 2.57$\pm$0.18\\
\enddata
\tablecomments{Numbers without errors are fixed to the quoted value }
\tablenotetext{a}{fixed}
\label{tab:lines}
\end{deluxetable*}

In Fig. \ref{fig:nolines}, we show the \chandra HETG spectrum of HD110432. Below (in red), we have plotted the HETG spectrum of $\gamma$ Cas analyzed in Smith et al. (2004), in the same wavelength interval, scaled and offset to allow comparison. In general terms the spectra of both objects are very similar. There are, however, some noticeable differences. Evidently, the Fe \textsc{xxvi} line is much broader in HD110432 than in $\gamma$ Cas. The Na \textsc{xi} is much stronger in HD110432. Despite the lower signal-to-noise ratio of the HD110432 spectrum, it seems evident that the Ne lines at 12.13 and 13.5 \AA\ respectively are much stronger in $\gamma$ Cas. The same seems to be true fo the Fe \textsc{xvii} lines around 15 \AA. 

The spectrum of HD110432 shows a wealth of emission lines available for plasma diagnostics, including the Fe K$\alpha$ fluorescence at 1.94 \AA. To extract their individual parameters, every line was fitted with Gaussians upon a three bremsstrahlung continuum, following the analysis made in the previous sections. Some lines remain unresolved. In such cases we fixed the line width to $\sigma=0.005$\AA. The measured $EW$, though, were found to be independent of this value. All identified and analyzed lines are listed in Table \ref{tab:lines}. 




In the sixth column we quote the temperature at which the emissivity of the ion is maximum ($kT_{\rm max}$), taken from the \textsc{atomdb}\footnote{\texttt{http://cxc.harvard.edu/atomdb}} data base. This temperature, however, does not reflect necessarily the temperature of the emitting plasma.


In the last column we quote the emission measure for each ion. This has been computed from the equation:

\begin{equation}
EM_{i}=\frac{4\pi d^{2} F_{i}}{\epsilon_{i}}
\end{equation}

where $F_{i}$ is the measured line flux, $\epsilon_{i}$ is the line emissivity at $T_{\rm max}$ and $d$ the distance to the star, which we take as $d=301$ pc (Hipparcos; Perryman 1997). As can be seen, values of $EM_{\rm i}$ are of the order of $10^{53-55}$ cm$^{-3}$. As stated before, the $kT_{\rm max}$ might not reflect the true temperature of the plasma and, therefore, the emissivity can have a lower value.

\begin{figure}
\includegraphics[angle=-90,width=\columnwidth]{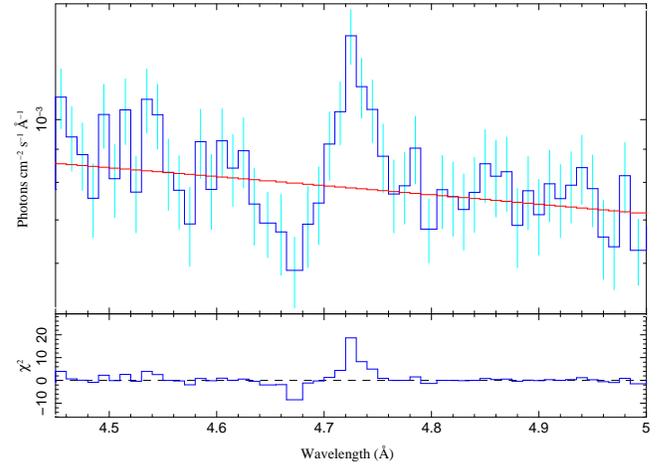}
\caption{The S \textsc{xvi} 4.73 \AA\ line presenting a possible P Cygni profile. }
\label{fig:pcygni}
\end{figure}

The S \textsc{xvi} 4.73 \AA\ line presents a possible P Cygni profile (Fig. \ref{fig:pcygni}). This possibility is rather interesting since it would point to the stellar wind as one of the possible sites for X-ray production in the system. P Cygni profiles have been observed by \chandra in the High Mass X-ray Binaries Vela X-1 and Cyg X-3 (Vilhu et al. 2009). The blue absorption to red emission peak separation implies a velocity of $\sim 3400$ km/s. Such high values have only been found in the terminal wind velocity $v_{\infty}$ of very early type stars like the O3 star HD 93129A \citep{taresch97}. This is, however, much larger than the wind velocity measured through the UV C\textsc{iv} lines \citep{smith06} in HD110432 and would rather suggest some violent localised outflow.  Unfortunately, no other line presents such a profile in our spectrum and further monitoring will be required to address this issue.

\subsubsection{He like triplets}

\begin{figure}
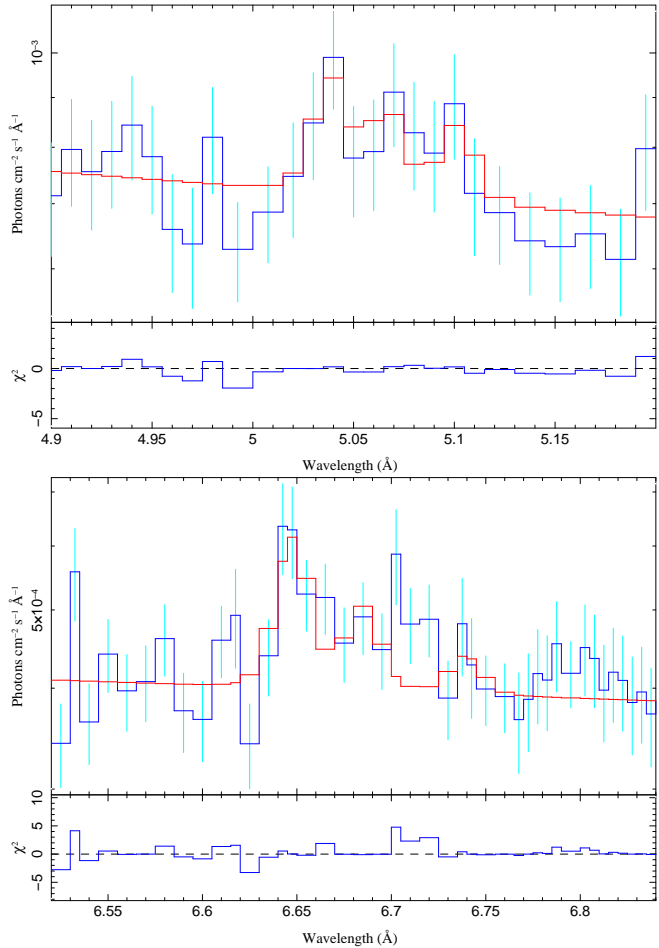

\includegraphics[angle=-90,width=\columnwidth]{sxv.ps}
\includegraphics[angle=-90,width=\columnwidth]{sixiii.ps}
\caption{Top: S \textsc{xv} triplet. The $r$, $i$ and $f$ components are clearly discerned. The positions and widths of the gaussians have been fixed while the normalization is left free.  Bottom: Si \textsc{xiii} triplet. The excess residuals at 6.72 \AA\ are roughly consistent with Mg \textsc{xii} Ly$\gamma$ because Mg \textsc{xii} Ly$\alpha$ is strong.}
\label{fig:triplets}
\end{figure}

Three He like triplets have been identified in the spectrum. Although the signal-to-noise ratio is low (see Fig. \ref{fig:triplets}) they can provide useful constrains as temperature and density diagnostics for the plasma \citep{porquet00}. The three lines visible in the triplets are called $r$, $i$ and $f$ which stand for the {\it resonance}, {\it intercombination} and {\it forbidden} transitions respectively. The models used in the previous section do not fit perfectly all the triplets {\it individually}. For example, the {\it bvapec} composite model is able to describe the $r$ and $f$ transitions of both S \textsc{xv} and Si \textsc{xiii} (although slightly underestimated) but does not show any detectable $i$ transition. The {\it vmekal}, in turn, describes well the $r$, $i$ and $f$ transitions in the Si \textsc{xiii} (albeit, again, underestimating all three lines) while they are undetectable in the S \textsc{xv} triplet. Therefore, we use the approach described previously where a gaussian has been fit to every transition (when present) over a multiple bremsstrahlung continuum.

Using values in Table \ref{tab:results} for the Si \textsc{xiii} triplet we can compute the index $G=(i+f)/r=0.73$ where $r$, $i$ and $f$ stand for the fluxes of the corresponding lines in the triplet, respectively. This value is indicative of a high plasma temperature, of the order of 10 MK. This is also supported by the presence of Mg \textsc{xi} which is not too prominent if $T< 10^{7}$ K. On the other hand, the index $R=f/i=0.42$ points to a high plasma density of the order of $n_{e}\approx 10^{13-14}$ cm$^{-3}$. The high density is also supported by the absence of the $f$ component of the Mg \textsc{xi} triplet which tends to be suppressed at high densities. This value is consistent with the high end range deduced by \cite{smith98} for the hot plasma in \gcas, namely $n_{e}\approx 10^{11-13}$ cm$^{-3}$. 

A strong UV radiation field can mimic a high density plasma because the {\it forbidden } line decreases due to the depopulation of the upper ($^{3}S$) levels via photo-excitation while the {\it intercombination} line is increased when the lower ($^{3}P$) levels are populated with these electrons \citep{porquet01}. Consequently we have considered the the possible influence of the intense UV radiation from the B0.5III star photosphere of blackbody temperature $T=33000$ K diluted by a factor $W$. For that purpose we shall use also the $R$ value for the S \textsc{xv} triplet, $R=f/i=0.86$. \cite{blum72} have calculated the low density, low UV irradiation limit of this quantity, $R_{0}$, for Si \textsc{xiii}, 2.51, and for S \textsc{xv}, 2.04. The observed $R$ value will depart from these values according to the equation

\begin{displaymath}
R=\frac{R_{0}}{1+\frac{n_{e}}{n_{c}}+2W\frac{\phi}{\phi_{c}}}
\label{eq:r}
\end{displaymath}

where $\phi$ is the photoexcitation rate and $n_{c}$ and $\phi_{c}$ are the corresponding critical values for plasma density and photoexcitation rate respectively and $W$ is the geometrical dilution factor. We see that $R\rightarrow R_{0}$ when $n_{e}\ll n_{c}$ (low density limit) and $\phi \ll \phi_{c}$ (low UV irradiation limit). Consequently, a value of $R$ lower than (but not orders of magnitude lower than) $R_{0}$ means high densities and/or high UV irradiation of the X-ray emitting plasma. In order to gain some insight on the possible plasma emission location we shall use the values of $\phi/\phi_{c}$ tabulated by \cite{blum72} for a blackbody of $10^{5}$ K and dilution factor $W=0.5$ (stellar surface) scaled here to the photospheric temperature $T=33000$ K and arbitrary $W$. We obtain $\phi/\phi_{c}=74$ for Si \textsc{xiii} and $\phi/\phi_{c}=17$ for S \textsc{xv}. We shall assume, further, an average plasma density for HD110432 of the order of $n_{e}\approx 10^{13}$ cm$^{-3}$, consistent with the observed $R$ value for HD110432 and that derived for \gcas. In that case the $n_{e}/n_{c}\approx 1$ for Si \textsc{xiii} (for which $n_{\rm c}=4\times 10^{13}$) while for S \textsc{xv} it will be $n_{e}/n_{c}\approx 0.1$ ($n_{\rm c}=1.9\times 10^{14}$). With these values we obtain dilution factors of 0.03 and 0.04 for Si \textsc{xiii} and S \textsc{xv} respectively. Now, recalling that $W=(1/2)[ 1- ( 1- (R_{*}/r))^{1/2}]$, where $r$ is the distance from the X-ray plasma site to the stellar center, we can locate the plasma at a distance of $r\simeq (2-3)R_{*}$. Although the associated errors are large ($\sim 50\%$) the small values of $R$ with respect the low density, low UV flux limit make it clear that neither the density nor the irradiation terms can be neglected in Eq. \ref{eq:r} locating the emission sites relatively close to the photosphere of the Be star at $r<3R_{*}$.   

Alternatively, this intense UV radiation field could come from the radiation of a putative WD companion. For the sake of comparison, we will take the WD surface temperature of the order of $10^{5}$ K. In such a case, this will be the primary ionizing field and we can use directly the values of $\phi/\phi_{c}$ calculated by \cite{blum72} to obtain dilution factors of $W=0.76$ for Si\textsc{xiii} and $W=0.2$ for S \textsc{xv}. These would translate into emission regions located at $r\sim [1.4-1.5]R_{WD}$ respectively, in units of the (putative) WD radius. That is to say, the X-ray emission would be produced very close to the WD surface, as might be expected if the X-rays are produced via accretion.





\subsubsection{The Fe complex}


In the first panel of Figures \ref{fig:vmekal} and \ref{fig:bvapec}, we plot the \chandra HETG spectrum of the Fe complex modeled with three \texttt{vmekal} and \texttt{bvapec} components of different temperatures and absorption columns, respectively. Clearly, the highly ionized species are significantly broader than the line from neutral Fe. While the width of the latter is not resolved in the \chandra spectrum, the former show a $FWHM \approx 0.02$ \AA\ (see Table \ref{tab:results}). Within the \texttt{vmekal} model, no additional correction has been applied. In the \texttt{bvapec} model we modify the resulting spectrum with some turbulence velocity $v_{turb}$. The hot component requires a $v_{turb}\approx 1200$ km/s which can account perfectly for the broadening of these lines. Consequently, the warm and cool components do not require any turbulent broadening. 

The Fe \textsc{xxv} line seems to present a double peak with $\Delta\lambda=0.009$ \AA. The circumstellar envelope in HD110432 is, contrary to the case of \gcas, seen almost edge on. As a result, its optical spectrum shows clearly double peaked Fe\textsc{ii} emission lines (LO07). In the Fe \textsc{xxv} line, the $\Delta \lambda$ corresponds to a separation of $\sim 1500$ km/s which is much larger than the typical Keplerian velocity of the disk but of the order of the wind velocity.

The $EW$ of the Fe \ka line is 89.78 eV. Using the curve of growth obtained by \cite{torr10} we get a density columns of the reprocessing material $N_{\rm H} = 3\times 10^{24}$ cm$^{-2}$. In contrast, the H density column obtained by \cite{smith06} from optical spectra is $3^{+2}_{-1}\times 10^{22}$ cm$^{-2}$. Fluorescence emission is also seen from Si \ka\ indicating the presence of large reservoirs of cold neutral material in the star or reflection off the (hypothetical) WD surface.


\section{Discussion}

\begin{figure}
\includegraphics[angle=0,width=\columnwidth]{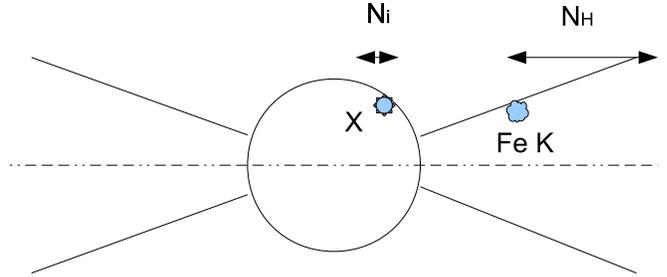}
\caption{Sketch showing the possible location of an X-ray source, above the disk, based on Fig. 4 of Smith et al. (2004). The source has an \emph{emission} column  $N_{\rm i}$ and illuminates the disk producing fluorescence. Both, ion emission, \fe \ka and continuum are absorbed by the same amount $N_{\rm H}$ from the circumstellar disk. The higher the lattitude, the smaller the absoprtion. The observer is located to the right.}
\label{fig:sketch_1}
\end{figure}

In general we see that, despite its possible peculiarities \citep{lopes10}, the spectrum of HD110432 seems to be reminiscent of \gcas (see for example, Smith et al. 2004, Fig. 2) strengthening the definition of the class. Consistently with LO07, the correct description of the spectrum requires at least three different thermal components of different temperatures and, as shown in this work, three different absorption columns. A hard tail is definitely present in HD110432 extending up to, at least, 50 keV. However, the nature of the hard component has not been unambiguously established for it can be equally well described by a thermal component or a powerlaw. A powerlaw could be a signature of synchrotron emission. 


The absorption columns deduced from the fits to the X-ray continuum range from $N^{\rm abs}_{\rm H}\sim 0.15\times 10^{22}$ cm$^{-2}$ to $3\times 10^{22}$ cm$^{-2}$. The lower value is compatible with the ISM medium ($0.2\times 10^{22}$ cm$^{-2}$, Ranchford et al. 2001). The high end value is consistent with the column deduced from the optical data for the envelope, namely $N^{\rm opt}_{\rm H}=3^{+2}_{-1}\times 10^{22}$ cm$^{-2}$ \citep{smith06}.



\subsection{X-ray active star or accretion onto a WD?}

The available data represents an isolated instance in the life of HD110432 and is insufficient, by itself, to settle this controversy. 
In principle, the described situation could be produced either by X-ray centers over the Be star photosphere, in the active star disk interaction paradigm or by accretion over a compact object orbiting relatively close to the stellar surface. Notwithstanding, we have presented here several pieces of evidence that can provide valuable new light onto the nature of the system when joined with previous results.

There is some evidence that the source is persistent. Indeed it has been clearly detected by all the X-ray telescopes that have pointed at it: HEAO, {\it BeppoSAX}, {\it RXTE}, {\it XMM}, \chandra and \suzaku. None of these observations was performed in any particular time or phase, yet the source was clearly detected and the lightcurve showed no eclipses. Unfortunately those observations do not span in time enough to firmly conclude that the source is non-eclipsing. In order to check this hypothesis we have analyzed the {\it RXTE} - All Sky Monitor (ASM) lighcurve for HD110432. The source is in the very limit of detection of ASM and most of the bins are compatible with zero flux. A continuus coverage with a large area collector all sky monitor would be important to address this point. If we assume that the source is persistent and non-eclipsing, either the X-ray emission is uniformly distributed over the area or the inclination of the system is such that the putative degenerate companion is never eclipsed. 

As we have seen, the available data could favor an origin of the X-ray emission close to the stellar surface, at distances $r<3R_{*}$ from the center of the Be star ($1R_{*}$ is the stellar surface). In such a case the compact object should be placed in a orbit with radius $r<3R_{*}$ with an orbital period between 4 and 8 days. Given that we do not observe periodic outbursts at these periods the orbit should be coplanar with the circumstellar disk. In order not to observe any eclipse the inclination of the system should be such that $\cos i > R_{*}/r$ which delivers the constrain $i < 60 - 70^{\rm o}$ for orbital radii of $2-3R_{*}$ respectively. This is very unlikely. The optical spectrum of HD110432 shows all its lines in emission with double peaks (see LO07, Fig. 1, and Smith \& Balona 2006) clearly indicating that the system is seen with $i\approx 90^{\rm o}$.  

On the other hand, as we have seen in the previous section, we can not exclude that the X-ray emission is produced very close to the stellar surface $r\sim 1.5R_{WD}$ of the putative WD companion. The multiplicity of absorbing columns is, however, difficult to reconcile with a single X-ray source embedded into the circumstellar disk. In any case, at least part of the X-ray emission appears to be absorbed by a column which is compatible with the density column of the circumstellar disk. Since the system is seen edge on, this would be in agreement with an X-ray source located close to the Be stellar surface which, in turn, argues in favor of the Be star photosphere being the main source of ionizing UV photons. The continuum monitoring of the source to detect or rule out the presence of eclipses would be, thus, important in understanding the nature of HD110432.

On the other hand, an X-ray emission distributed on or over the surface of the star could account naturally for the different absorbing density columns for each component as depicted in Fig \ref{fig:sketch_1}. Since the system is seen close to edge on and the disk flares outwards (in order to be in hydrostatic equilibrium) the lower the latitude of the X-ray source the higher the absorption. Thus, a number of small X-ray sources distributed on or above the stellar photosphere at several latitudes would account naturally for the characteristics of the observed emission.

The lightcurve shows a high variability and flaring but all efforts made so far have failed to detect any coherent pulsation. The persistent lack of detection of any coherent modulation strongly argues against the degenerate companion scenario. In conclusion, with the available data at hand the X-ray active standalone star seems to emerge as the least problematic explanation. The mechanism behind such energetic emission, however, remains a mystery and is challenging. In particular, the extreme temperatures of the order of 15-30 keV or, alternatively, the presence of non thermal components is not explained under this paradigm. The fact that other isolated Be stars with similar spectral types and comparably large circumstellar disks do not show any X-ray emission at all, is not explained either. Further observations will be required to establish the nature of this mysterious object. In particular, it would be necessary to: a) obtain a high signal-to-noise ratio spectrum of the high energy tail in order to unambiguously establish the nature of the hard component using, for example, NuSTAR. This would require, probably, the \emph{simultaneous} use of \chandra and/or {\it XMM-Newton}. b) Monitor the star with an all sky monitor to establish or rule out the persistence of the X-ray emission and the presence or lack of eclipses (with MAXI or LOFT), c) a monitoring campaign with a large collecting area telescope (to reduce the long exposure times) of the triplets to look for some periodic modulation that could arise from the (co)rotation of the X-ray center(s) around the star.

\acknowledgments

The authors would like to acknowledge Myron Smith for his permission to use their original HETG $\gamma$ Cas data on Fig. \ref{fig:nolines}. This work has been supported by the Spanish
Ministerio de Ciencia, Tecnolog\'{\i}a e Innovaci\'on (MCINN) through the grant AYA2010-15431. JMT acknowledges the use of the computer facilities made available through the grant AIB2010DE-00057.  


\end{document}